\documentclass[twocolumn,showpacs,amsmath,amssymb]{revtex4}
\usepackage{amssymb}
\usepackage[dvips]{graphicx}
\begin{document}

\title{Superconducting Gap, Normal State Pseudogap and Tunnelling Spectra of Bosonic and
Cuprate Superconductors}
\author{A. S. Alexandrov and J. Beanland}

\affiliation{Department of Physics, Loughborough University,
Loughborough LE11 3TU, United Kingdom\\}

\begin{abstract}
We develop a theory of normal-metal - superconductor (NS) and superconductor - superconductor (SS) tunnelling in \textquotedblleft bosonic"
superconductors with strong attractive correlations taking into account coherence effects in single-particle excitation spectrum and
disorder. The theory accounts for the existence of two  energy scales, their temperature and doping dependencies, asymmetry and
inhomogeneity of tunnelling spectra of underdoped cuprate superconductors.

\end{abstract}


\maketitle

 Soon after  the discovery of high-T$_c$ superconductivity \cite{muller},
a number of tunnelling, photoemission, optical, nuclear spin
relaxation and electron-energy-loss spectroscopies discovered an
anomalous large gap in cuprate superconductors  existing well above
the superconducting critical temperature, T$_c$. The gap, now known
as the pseudogap, was originally assigned \cite{aleray} to the
binding energy of real-space preformed hole pairs - small bipolarons
- bound by a strong electron-phonon interaction (EPI).  Since then,
alternative explanations of the pseudogap have been proposed,
including preformed Cooper pairs \cite{emery}, inhomogeneous
charge distributions containing  hole-rich and  hole-poor domains \cite{demello},
 or competing quantum phase transitions \cite{shen2007}.

Present-day scanning tunnelling  (STS) \cite{gomes,davis2009,kato},
intrinsic tunnelling \cite{krasnov} and angle-resolved photoemission
(ARPES) \cite{shen2007} spectroscopies have offered a tremendous
advance into the understanding of the pseudogap phenomenon in
cuprates and some related compounds. Both extrinsic (see
\cite{gomes,kato} and references therein) and intrinsic
\cite{krasnov} tunnelling as well as high-resolution ARPES
\cite{shen2007} have  found another energy scale, reminiscent of a
BCS-like \textquotedblleft superconducting" gap that opens at T$_c$
accompanied by the appearance of Bogoliubov-like quasi-particles
\cite{shen2007} around the node. Earlier experiments with a
time-resolved pump-probe demonstrated two distinct gaps, one  a
temperature independent pseudogap and the other a BCS-like gap
\cite{dem}. Also, Andreev reflection experiments revealed a much
smaller gap edge than the bias at the tunnelling conductance maxima
in a few underdoped cuprates \cite{deutscher}. Another remarkable
observation is the spatial nanoscale inhomogeneity of the pseudogap
observed with STS \cite{gomes,davis2009,kato} and presumably related
to an unavoidable disorder in doped cuprates, Fig.1a. Essentially,
the doping and magnetic field dependence of the superconducting gap
compared with the pseudogap and their different real space  profiles
have prompted an opinion that the pseudogap is detrimental to
superconductivity and connected to a quantum critical point rather
than to preformed Cooper pairs \cite{krasnov}.
\begin{figure}
\begin{center}
\includegraphics[angle=-90,width=0.56\textwidth]{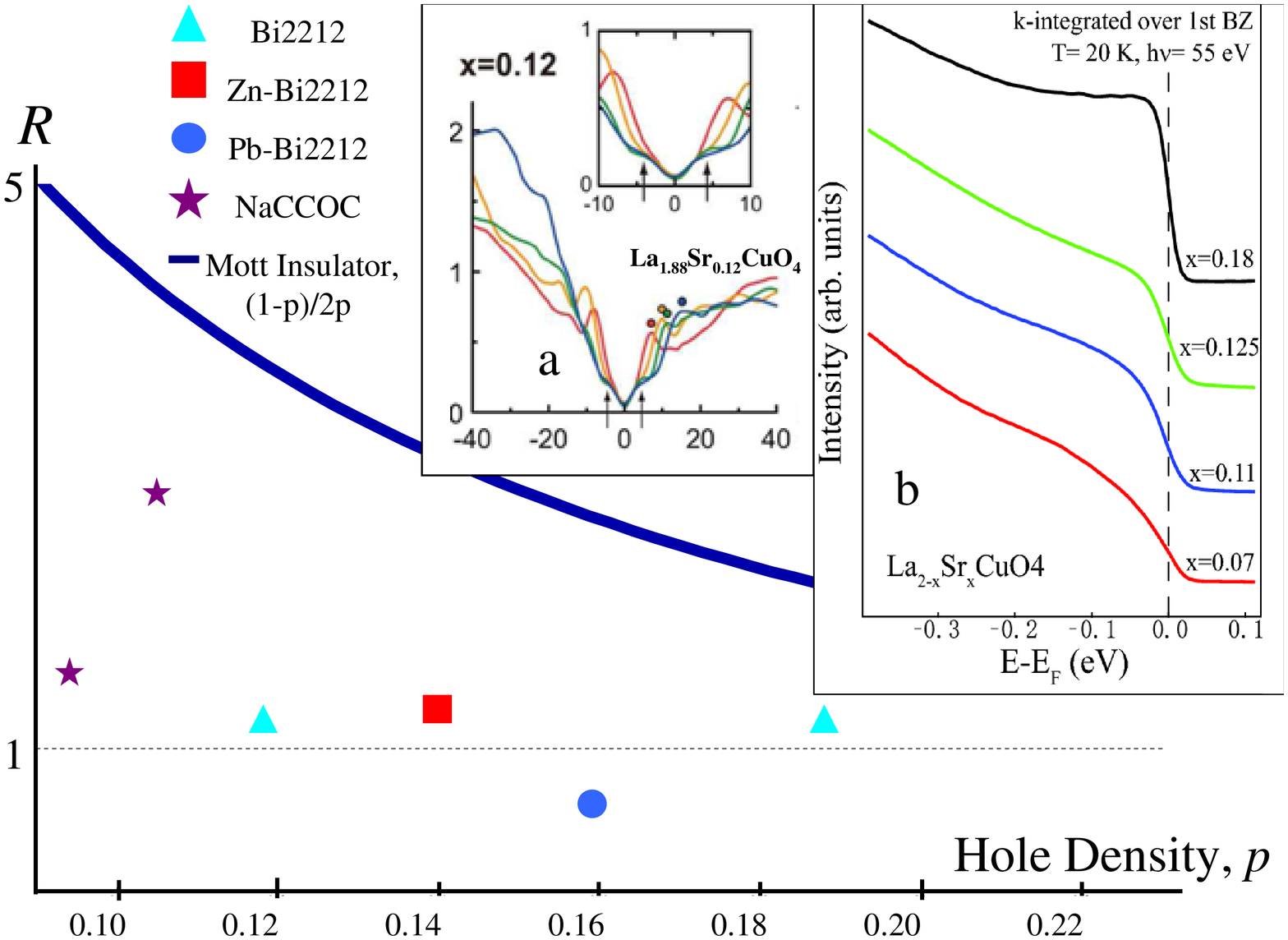}
\vskip -0.5mm \caption{(Colour online) Ratio,
$R=I_{ns}(-100)/I_{ns}(100)$, of the negative-bias NS tunnelling
conductance to the positive bias conductance \cite{asymmetry}
integrated from $0$ meV  to $\mp 100$ meV, respectively, for a few
cuprate superconductors in a wide range of  atomic hole density,
$p$. Inset ($a$): atomically resolved asymmetric STS spectra of
La$_{1.88}$Sr$_{0.12}$CuO$_{4}$ at 4.2K acquired at different points
of the scan area \cite{kato}; ($b$): momentum-integrated
photoemission showing no vHs of DOS \cite{private}. }
\end{center}
\end{figure}
Nevertheless without a detailed microscopic theory that can describe
highly unusual tunnelling and ARPES spectra, the relationship
between the pseudogap and the superconducting gap remains a mystery
\cite{shen2007}. It has become increasingly likely that NS and SS
tunnelling spectra of underdoped cuprates do not agree with the
simplest BCS spectra neither s-wave nor d-wave. Apart from the
almost temperature independent pseudogap, $\Delta_p$ with
$2\Delta_p/k_BT_c$ often many times larger than the BCS ratio
($\approx 3.5)$, there is an  asymmetry in NS tunnelling, Fig.1. The
integrated conductance for the negative bias is  larger than for the
positive bias in many samples. The  van Hove singularity (vHs) of
the density of states (DOS) is ruled out as a possible origin of the
asymmetry since it is absent in the momentum-integrated
photoemission, Fig.1b. The asymmetry is expected for conventional
semiconductors or Mott-Hubbard insulators
($I_{ns}(-\infty)/I_{ns}(+\infty)=(1-p)/2p$), but neither of them
account for its magnitude, Fig.1, if disorder and matrix elements
are not considered. Here we develop the theory of NS and SS
tunnelling in bosonic superconductors with strong attractive
correlations \cite{alebook} by taking into account disorder and coherence effects
in a single-particle excitation spectrum. Our theory accounts for
peculiarities in extrinsic and intrinsic tunnelling in underdoped
cuprate superconductors.

Recent  Monte Carlo  calculations show
that high-T$_c$ superconductivity  cannot be explained by
the simplest repulsive Hubbard  model \cite{imada}, so that one has to extend the model to get some superconducting order \cite{baer}.  On the other hand  even a moderate
EPI significantly increases the superconducting condensation energy  \cite{hardy} stabilizing mobile small bipolarons  \cite{hague,bonca}, as anticipated for strongly correlated
electrons in highly polarizable ionic lattices \cite{alebook}.
Real-space pairs, whatever the pairing interaction is, can be
described as a charged Bose liquid on a lattice, if the carrier
density is relatively small \cite{alebook}.
The superfluid state of such a liquid is the true Bose-Einstein
condensate (BEC), rather than a coherent state of overlapping Cooper
pairs. Single-particle excitations of the liquid are thermally
excited single polarons propagating in a doped insulator band or are
localised by impurities. Different from the BCS case, their {\it
negative} chemical potential, $\mu$, is found outside the band by
about half of the bipolaron binding energy, $\Delta _{p}$, both in
the superconducting and normal states \cite{alebook}. Here, in the
superconducting state (T$<$T$_{c}$), following Ref.\cite{aleand} we
take into account that polarons interact with the condensate via the same
potential that binds the carriers, so that the single-particle
Hamiltonian in the Bogoliubov approximation is
\begin{equation}
H_0= \sum_\nu \left [\xi_\nu p_{\nu}^\dagger
p_\nu+\left({1\over{2}}\Delta_{c\nu}p_{\bar{\nu}}^\dagger p_{\nu}
^\dagger + H.c.\right)\right],
\end{equation}
where $\xi_\nu= E_\nu -\mu$, $E_\nu$ is the normal-state
single-polaron energy spectrum in the crystal field and disorder
potentials renormalised by EPI and spin-fluctuations, and
$\Delta_{c\nu}=-\Delta_{c \bar{\nu}}$ is the coherent potential
proportional to the square root of the condensate density, $\Delta
_{c}\propto \sqrt{n_{c}(T)}$. The operators $p_{\nu}^\dagger$ and
$p_{\bar{\nu}}^\dagger$ create a polaron in the single-particle
quantum state $\nu$ and in the time-reversed state $\bar{\nu}$,
respectively. As in the BCS case the single quasi-particle energy
spectrum, $\epsilon_{\nu}$, is found using the Bogoliubov
transformation, $p_\nu=u_\nu \alpha_\nu+v_\nu\beta_\nu^{\dagger}$,
$p_{\bar {\nu}}=u_\nu \beta_\nu -v_\nu\alpha_\nu^{\dagger}$,
$\epsilon_{\nu}=\left[ \xi_{\nu}^2+\Delta _{c \nu}^{2}\right]
^{1/2}$, with $u_{\nu}^2,\ v_{\nu}^2=(1\pm\xi_\nu/\epsilon_\nu)/2$.
This spectrum is  different from the BCS quasi-particles because the
chemical potential is negative with respect to the bottom of the
single-particle band, $\mu =-\Delta _{p}$. A single-particle gap,
$\Delta$, is defined as the minimum of $\epsilon_{\nu}$.  Without
disorder, for a point-like pairing potential with the s-wave
coherent gap, $\Delta_{c {\bf k}}\approx \Delta_{c}$, one has
\cite{aleand} $\Delta(T)=\left[ \Delta _{p}^{2}+\Delta
_{c}(T)^{2}\right] ^{1/2}$. The full gap varies with temperature
from $\Delta (0)=\left[ \Delta_{p}^{2}+\Delta _{c}(0)^{2}\right]
^{1/2}$ at zero temperature down to the temperature independent
$\Delta=\Delta _{p}$ above T$_{c}$, which qualitatively describes
some earlier  and more recent \cite{krasnov} observations including
the Andreev reflection in cuprates (see \cite{aleand} and references
therein).
\begin{figure}
\begin{center}
\includegraphics[angle=-90,width=0.40\textwidth]{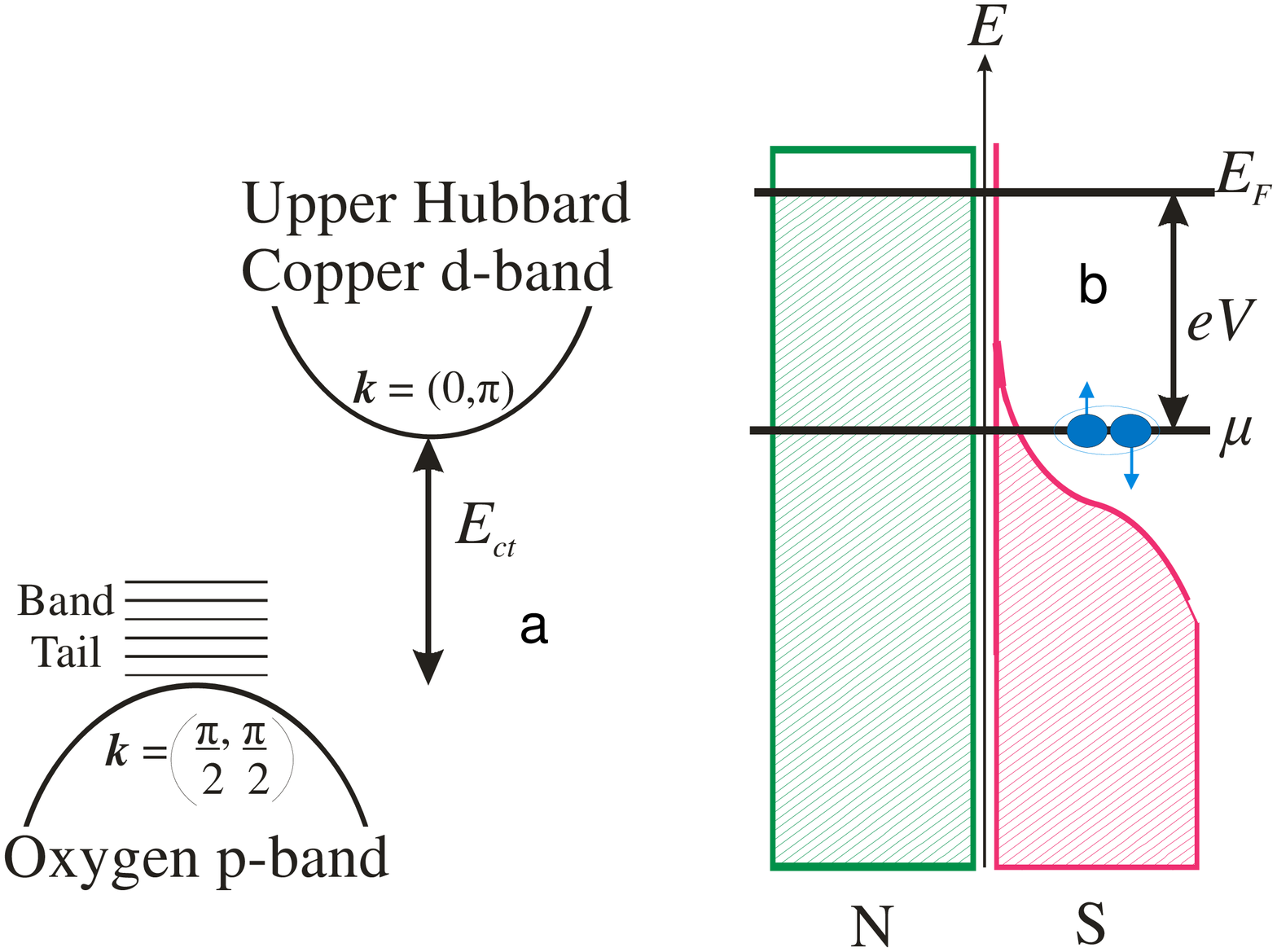}
\vskip -0.5mm \caption{(Colour online) LDA+GTB energy band structure
of underdoped cuprates with the impurity localised states
\cite{alekim} shown as horizontal lines in ($a$); NS model densities
of states $(b)$ showing  the bandtail  in the bosonic
superconductor.}
\end{center}
\end{figure}
The NS and SS tunnelling transitions are described with the
tunnelling Hamiltonians, $H_{ns}$ and $H_{ss}$ respectively,
\cite{aletun}, which are perturbations:
\begin{eqnarray}
&&H_{ns}=P\sum_{\nu \nu'}p_{\nu'}^\dagger c_\nu+ BN^{-1/2} \sum_{\nu
\nu' \eta'}b_{\eta'}^\dagger p_{\bar{\nu'}} c_\nu + \text{h.c.}, \cr
&&H_{ss}= P\sum_{\nu, \nu'}p_{\nu'}^\dagger p_\nu+ BN^{-1/2}\times
\cr && \sum_{\nu \nu' }\left( \sum_{\eta} p_{\nu'}^\dagger
p_{\bar{\nu}}^\dagger
 b_{\eta} +\sum_{\eta'}b_{\eta'}^\dagger
p_{\nu} p_{\bar{\nu'}}\right) + \text{h.c.}. \label{ss}
\end{eqnarray}
Here $c_{\nu},\ p_{\nu'}$ and $b_{\eta'}$ annihilate a carrier in
the normal metal, a single polaron and a composed  boson in the
bosonic superconductor respectively,  $N$ is the number of unit
cells. Generally the tunnelling matrix elements with ($B$) and
without ($P$) involvement of the composed boson are different,
$B\gtrsim P$, because the presence of an additional hole lowers the
tunnelling barrier for an injection of the electron \cite{aletun}.
Applying the Bogoliubov transformation to Eqs.(\ref{ss}) and the
standard perturbation theory yields the following current-voltage
characteristics:
\begin{eqnarray}
&&I_{ns} (V)={2\pi e P^2\over{\hbar}} \sum_{\nu \nu'}[u_{\nu'}^2
(F_{\nu}-f_{\nu'}) \delta(\xi_\nu+eV-\epsilon_{\nu'}) +\cr &&
v_{\nu'}^2 (F_{\nu}+f_{\nu'}-1)\delta (\xi_\nu+eV+\epsilon_{\nu'})]
+ {2\pi e B^2\over{\hbar}}\sum_{\nu \nu'}  \cr && \left[ u_{\nu'}^2
[F_{\nu}f_{\nu'}-(x/2) (1-F_\nu-f_{\nu'})] \delta
(\xi_\nu+eV+\epsilon_{\nu'})+ \right.\cr && \left.v_{\nu'}^2
[F_{\nu}(1-f_{\nu'})+(x/2) (F_\nu-f_{\nu'})]\delta
(\xi_\nu+eV-\epsilon_{\nu'})\right], \nonumber \\ \label{Ins}
\end{eqnarray}
where $F_{\nu}=1/[\exp(\xi_{\nu}/k_BT)+1]$,
$f_{\nu'}=1/[\exp(\epsilon_{\nu'}/k_BT)+1]$ are
 distribution functions of carriers in the normal metal and single
quasi-particles respectively,  and  $x/2$ is the atomic density of
composed bosons in the superconductor,  and
\begin{eqnarray}
&&I_{ss} (V)={2\pi e P^2\over{\hbar}} \sum_{\nu \nu'}[(u_{\nu}^2
u_{\nu'}^2+v_{\nu}^2 v_{\nu'}^2) (f_{\nu}-f_{\nu'})\times \cr &&
\delta(\epsilon_\nu+eV-\epsilon_{\nu'})+ u_{\nu}^2v_{\nu'}^2
(f_{\nu}+f_{\nu'}-1)\times \cr && [\delta
(\epsilon_\nu+eV+\epsilon_{\nu'})-\delta
(\epsilon_\nu-eV+\epsilon_{\nu'})]]+\cr && {2\pi e B^2\over{\hbar}}
\sum_{\nu \nu' }[u_{\nu}^2u_{\nu'}^2
((1-f_\nu-f_{\nu'})x/2-f_{\nu}f_{\nu'})+ \cr && v_{\nu}^2v_{\nu'}^2
((1-f_\nu-f_{\nu'})x/2+(1-f_{\nu})(1-f_{\nu'}))]\cr &\times& [\delta
(\epsilon_\nu-eV+\epsilon_{\nu'})-\delta
(\epsilon_\nu+eV+\epsilon_{\nu'})]+ \cr && 2u_{\nu}^2v_{\nu'}^2
[(f_{\nu'}-f_{\nu})x/2-f_{\nu}(1-f_{\nu'})]\cr &\times& [\delta
(\epsilon_\nu+eV-\epsilon_{\nu'})-\delta
(\epsilon_\nu-eV-\epsilon_{\nu'})], \label{Iss}
\end{eqnarray}
where $V$ is the voltage drop across the junction.  For more transparency  we neglect the boson energy dispersion in Eqs.(\ref{Ins}, \ref{Iss}), assuming that   bosons are sufficiently  heavy, so  their
bandwidth is relatively small.  Here we adopt the
\textquotedblleft LDA+GTB" band structure with impurity bandtails
 near
$(\pi/2,\pi/2)$ of the Brillouin zone, Fig.2a, which explains the charge-transfer gap,
$E_{ct}$, the nodes and sharp \textquotedblleft quasi-particle" peaks,  and  the high-energy
\textquotedblleft waterfall" seen in ARPES  \cite{alekim}. The chemical potential is found in
the single-particle bandtail within the charge-transfer gap at the
bipolaron mobility edge, Fig.2b, in agreement with the SNS
tunnelling experiments \cite{boz}. Such a band structure explains an
insulating-like low temperature normal-state resistivity  as well as
many  other unusual properties of underdoped cuprates
\cite{alebook}. If the characteristic bandtail width of DOS, $\Gamma$, is
sufficiently large compared with the coherent gap,
$\Gamma\gtrsim\Delta_{c\nu}$, one can factorize the quasi-particle
DOS as $\rho(E)\equiv \sum_{\nu} \delta(E-\epsilon_\nu)\approx
[\rho_{n}(E)+\rho_n(-E)]\rho_s(E)$ for any symmetry of the coherent
gap. Here $\rho_{n}(E)$ is the normal state DOS of the doped
insulator with the band-tail, Fig.2a, and
$\rho_s(E)=E/\sqrt{E^2-\Delta_c^2}$ for the s-wave gap, or
$\rho_s(E)=(2/\pi)[\Theta(1 - E/\Delta_0)E K(E/\Delta_0)/\Delta_0 +
\Theta(E/\Delta_0 - 1)K(\Delta_0/E)]$ for a  d-wave gap,
$\Delta_{c\nu}= \Delta_0 \cos (2\phi)$ \cite{maki} ($K(x)$ is the
complete elliptic integral and $\phi$ is an angle along the constant
energy contour). One can neglect the energy dependence of the normal
metal DOS. Then differentiating Eq.(\ref{Ins}) over the voltage and
integrating yields the NS tunnelling conductance $\sigma_{ns}=
dI_{ns}/dV$ at zero temperature for the d-wave case,
\begin{eqnarray}
&&\sigma_{ns}\propto A^+\rho_s(|eV|)[\rho_n(-eV) + \rho_n(eV)]+\cr
&&A^-[1- 2\cos^{-1}(|eV|/\Delta_0) \Theta(1 - |eV|/\Delta_0)/\pi]\cr
&\times& [\rho_n(-eV) - \rho_n(eV)], \label{condm}
\end{eqnarray}
where $A^{\pm}=1\pm B^2\left[\Theta(-eV)+x/2\right]/P^2$,  and
$\Theta(E)$ is the Heaviside step function.
\begin{figure}
\begin{center}
\includegraphics[angle=-90,width=0.30\textwidth]{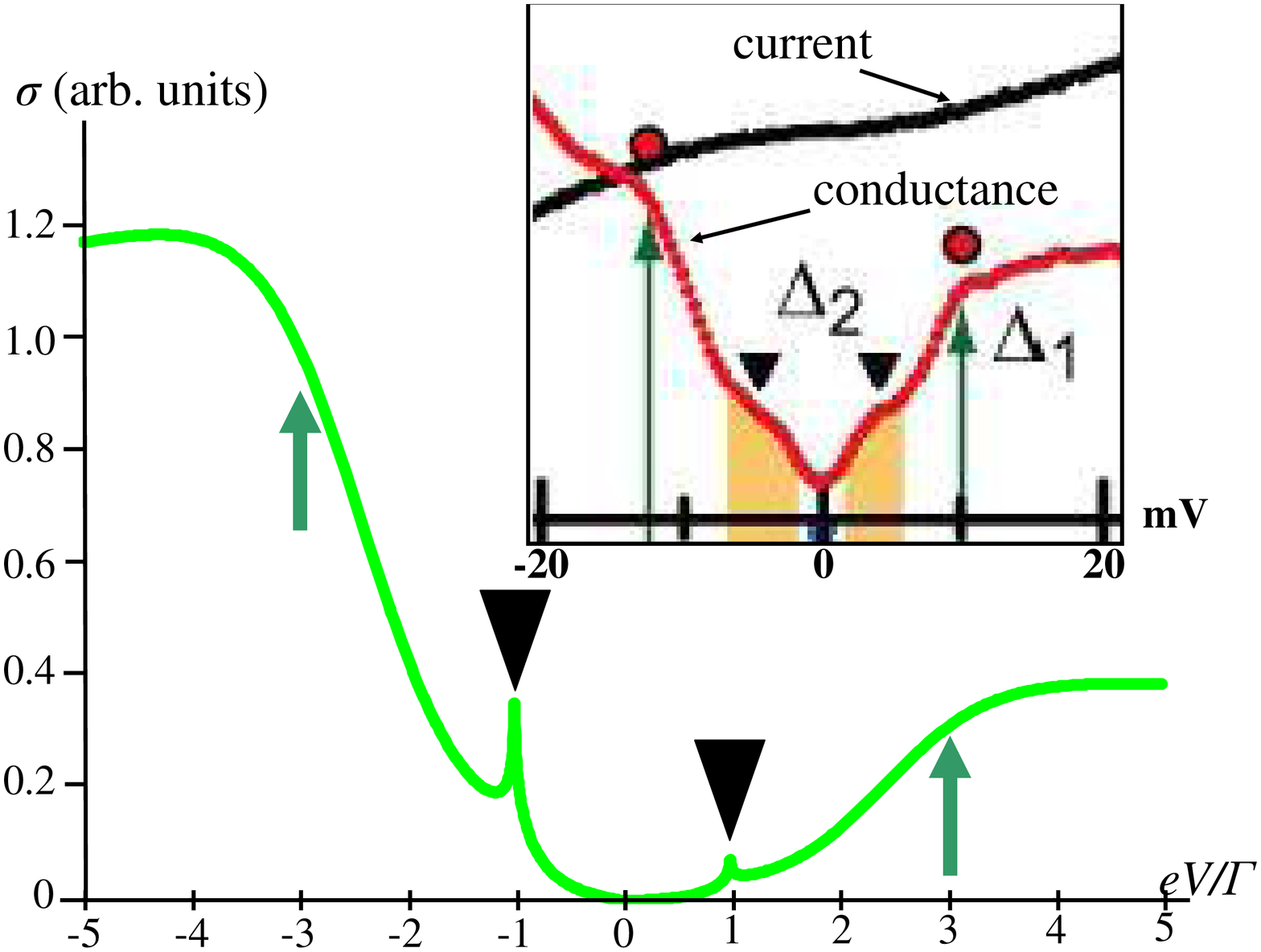}
\vskip -0.5mm \caption{(Colour online) Theoretical NS conductance,
Eq.(\ref{condm}), for $\Delta_0=\Gamma$,  $\Delta_p= 2.7 \Gamma$ and
$B=2.65 P$.  The   superconducting gap and the
pseudogap are shown with triangles and arrows, respectively. Inset shows a representative STS
spectrum of La$_{2-x}$Sr$_{x}$CuO$_{4}$ with $x=0.12$ at 4.2K
\cite{kato}. }
\end{center}
\end{figure}
The theoretical conductance, Eq.(\ref{condm}), calculated with a
model normal state DOS,
$\rho_n(E)/\rho_b=\{1+\tanh[(E-\Delta_p)/\Gamma]\}/2$, and the d-wave
superconducting DOS, $\rho_s(E)$,  is shown in Fig.3.  Our model $\rho_n(E)$ reflects the characteristic energy dependence of DOS in disordered doped insulators,
 which is a constant $\rho_b$ above the two-dimensional band
edge and an exponent deep in the tail. Any particular choice of $\rho_n(E)$ and the model
parameters can be made without affecting our conclusions
as long as the characteristic features are reflected in
this choice.  Eq.(\ref{condm}) captures all unusual signatures of the
experimental tunnelling conductance in underdoped cuprates, such as
the low energy coherent gap, the high-energy pseudogap, and the
asymmetry. In the case of atomically resolved STS one should replace
the averaged DOS $\rho_n(E)$ in Eq.(\ref{condm}) with a $local$
bandtail DOS $\rho_n(E, {\bf r})$, which depends on different points
of the scan area ${\bf r}$ due to a nonuniform dopant distribution.
As a result  the pseudogap shows nanoscale inhomogeneity, while the
low-energy coherent gap is spatially uniform, as observed
\cite{kato}, Fig.1a. Increasing doping level tends to diminish the
bipolaron binding energy, $\Delta_p$,  since the pairing potential
becomes weaker due to  a partial screening of EPI with low-frequency
phonons \cite{alekabmot}. However, the coherent gap, $\Delta_c$,
which is the product of the pairing potential and the square root of
the carrier density \cite{aleand}, can remain  about a constant or
even increase with doping, as also observed \cite{kato}.

In the case of the SS tunnelling we use Eq.(\ref{Iss}) to address
two unusual observations: a gapped conductance near and above T$_{c}$
and a negative excess resistance below T$_c$
\cite{krasnov,krasnovPRL}. Eq.(\ref{Iss}) is grossly simplified in
the normal state, where $\Delta_{c\nu}=0$,
\begin{eqnarray}
&&I_{ss} (V)=\frac{2\pi e P^2}{\hbar} \sum_{\nu\nu^{\prime}}(f_{\nu}-f_{\nu^{\prime}})\delta(\xi_\nu+eV-\xi_{\nu^{\prime}}) \cr
&&
+ \frac{2\pi e B^2}{\hbar} \sum_{\nu \nu^{\prime}}[(1-f_\nu-f_{\nu^{\prime}})x/2-f_{\nu}f_{\nu^{\prime}}) \cr
&&
\times[\delta(\xi_\nu-eV+\xi_{\nu^{\prime}})-\delta (\xi_\nu+eV+\xi_{\nu^{\prime}})].\label{IssN}
\end{eqnarray}
Near and above the transition but sufficiently below the pseudogap
temperature T$^*\equiv \Delta_p/k_B >$ T $\gtrsim$ T$_c$, and if the
voltage is high enough, $eV\gtrsim k_BT$, one can neglect
temperature effects in Eq.(\ref{IssN}) and approximate $f_{\nu}$
with the step function, $f_{\nu}=\Theta(-\xi_\nu)$. So using the
model normal state DOS yields
\begin{eqnarray}
&&I_{ss} (V)\propto \frac{a^2}{2(a^2-1)}\left[\ln
\frac{a^2(1+b^2)}{1+a^2b^2}-a^{-2}\ln \frac{a^2+b^2}{1+b^2}\right]+
\cr &&
\frac{B^2(1+x/2)}{P^2(a^2b^4-1)}\ln\frac{1+a^2b^2}{a(1+b^2)}+\frac{B^2xa^2}{2P^2(a^2-b^4)}\ln
\frac{a^2+b^2}{a(1+b^2)}, \nonumber \label{IssN2}\\
\end{eqnarray}
where $a=\exp(\left|eV\right|/\Gamma)$ and
$b=\exp(\Delta_p/\Gamma)$. When $b\gg 1$ and $x$ is not too small,
the  first two terms on the right hand-side of this equation are
negligible. Hence the tunnelling matrix elements and doping have
little effect on the shape of the current-voltage dependence.  At
sufficiently high voltages $eV\gtrsim k_BT$ the conductance (from
$\sigma(V)=dI_{ss}/dV$ with Eq.(\ref{IssN2})) accounts for the gaped
conductance in underdoped mesas of
Bi$_2$Sr$_2$CaCu$_2$O$_{8+\delta}$ near and above T$_c$, as shown in
Fig 4. The finite temperature neglected in Eq.(\ref{IssN2}) accounts
for some excess experimental conductance at low voltages in Fig.4
compared with the theoretical conductance.
\begin{figure}
\begin{center}
\includegraphics[angle=-90,width=0.33\textwidth]{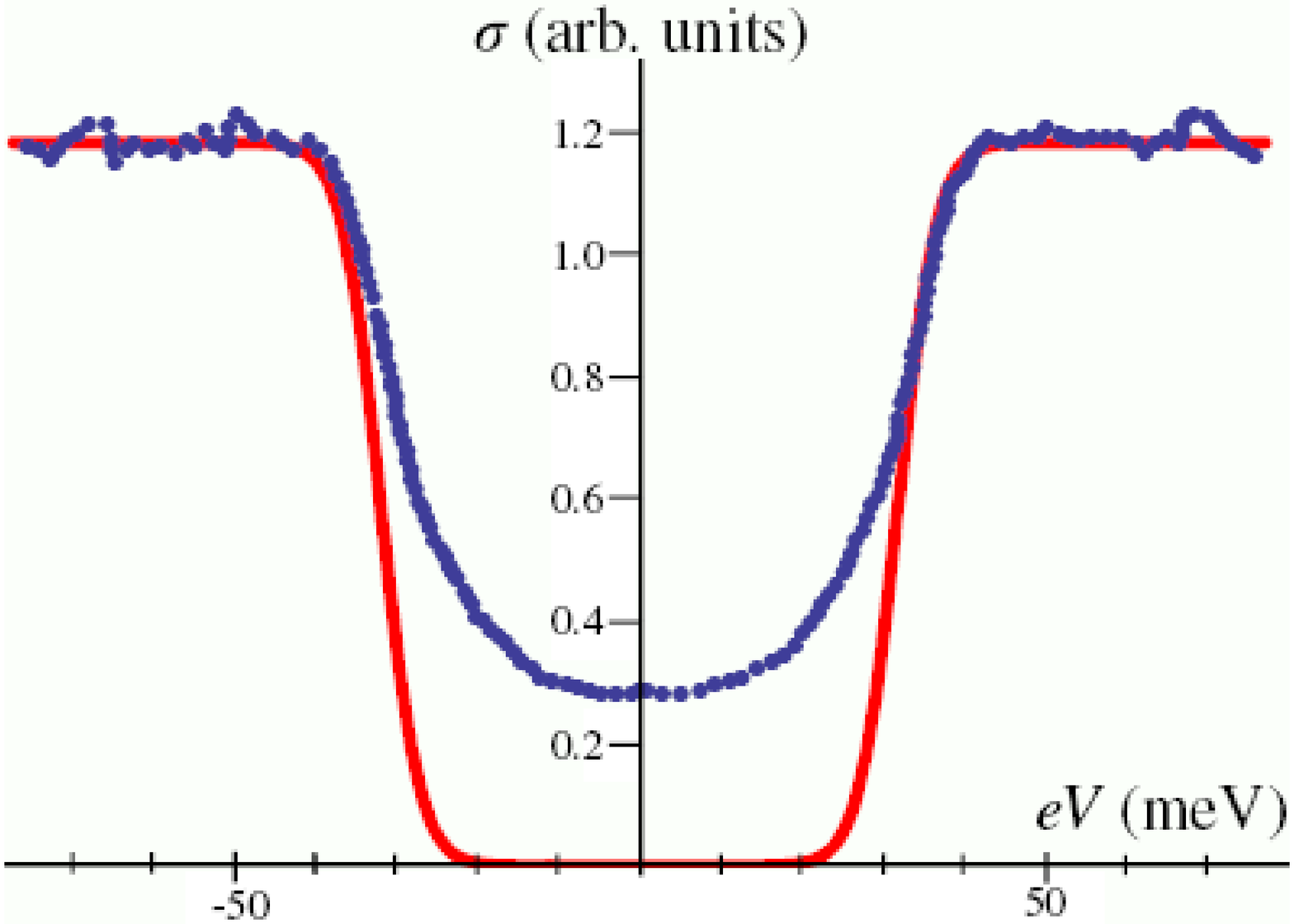}
\vskip -0.5mm \caption{(Colour online) Approximate normal state
tunnelling  conductance of  bosonic superconductor (solid line) with
$\Gamma=3.2$ meV and $\Delta_p=16$ meV compared with the
experimental conductance \cite{krasnov} (symbols) in mesas of
Bi$_2$Sr$_2$CaCu$_2$O$_{8+\delta}$ (T$_c$=95K) at $T=87$K.}
\end{center}
\end{figure}
The negative excess resistance below T$_c$ \cite{krasnovPRL} can be
explained by expanding Eqs.(\ref{Iss},\ref{IssN}) in powers of $eV$
giving a zero bias conductance. For low temperatures in the
superconducting state this is $\sigma_s (0) \propto T^{-1} \int
_{0}^{\infty} d\epsilon \rho_s(\epsilon)^2\cosh(\epsilon/2k_BT)^{-2}
$, and in the normal state $\sigma_n (0) \propto T^{-1} \int
_{-\infty}^{\infty} d\xi \rho_n(\xi)^2\cosh(\xi/2k_BT)^{-2}$.
 Estimating these integrals yields, respectively
$\sigma_s (0)\propto T^{-1} \exp(-\Delta_c/k_BT)$ for the s-wave
coherent gap, or $\sigma_s(0)\propto T^{2}$ for the d-wave gap, and
$\sigma_n (0)\propto T^{-1} \exp(-T^*/T)$. The latter expression is
in excellent agreement with the temperature dependence of the mesa
tunnelling conductance above T$_c$ \cite{krasnovPRL} (see also Ref.
\cite{alekabmot}). Extrapolating this expression to temperatures
below T$_c$ yields  the resistance ratio $R_s/R_n \propto
\exp[(\Delta_c/k_B-T^*)/T]$ (s-wave) or $R_s/R_n \propto
\exp(-T^*/T)/T^2$ (d-wave). Hence in underdoped cuprates, where $T^*
> \Delta_c/ k_B$, the zero-bias tunnelling resistance at
temperatures below T$_c$ is smaller than the normal state resistance
extrapolated from above T$_c$ to the same temperatures (i.e. the
negative excess resistance), as observed \cite{krasnovPRL}.

   In summary, we have developed the theory of tunnelling in bosonic
superconductors by taking into account coherence effects in the
single-quasi-particle energy spectrum, disorder and the realistic
band structure of doped insulators. The theory accounts for the
existence of two energy scales in the current-voltage NS and SS
tunnelling characteristics, their temperature and doping dependence,
and for the asymmetry and inhomogeneity of NS tunnelling spectra of
underdoped cuprate superconductors.

We are grateful  to Zhi-Xun Shen and Ruihua He for providing us with
their momentum-integrated ARPES data, Fig.1b and enlightening
comments. We greatly appreciate valuable discussions with Ivan
Bozovic, Kenjiro Gomes and Vladimir Krasnov and support of this work
by EPSRC (UK) (grant number EP/H004483).

\end{document}